\documentclass[12pt]{article}
\usepackage{graphicx}
\usepackage{amsmath}
\usepackage{amssymb}
\newcommand{\be}{\begin{equation}}
\newcommand{\ee}{\end{equation}}
\newcommand{\bea}{\begin{eqnarray}}
\newcommand{\eea}{\end{eqnarray}}
\newcommand{\pr}{\partial}

\newcommand{\bse}{\begin{subequations}}
\newcommand{\ese}{\end{subequations}}

\begin{document}
\title{Dynamic domain walls in Maxwell-dilaton background}
\author{Debaprasad Maity\footnote{E-mail: debu@imsc.res.in}\\
The Institute of Mathematical Sciences,\\ 
C.I.T. Campus, Tharamani,\\
Chennai - 600 113, India}
\maketitle
\begin{abstract}
Motivated by the well know
Chamblin-Reall solutions of $n$-dimensional background spacetime
in a dilaton gravity and the dynamics of a domain wall in the same backgrounds,
we have tried to generalize those solutions by 
including electromagnetic field in the bulk.
The electromagnetic field is assumed to be coupled 
with the scalar field in an exponential
way. Under the specific relations among the various
parameters in our model, 
we have found five different types of solutions. For every case, 
the solution has singularity. In these backgrounds, 
we have also studied the dynamics of domain wall. 
The energy densities which play the role of these interesting 
dynamics, are know to be
induced from the bulk fields through Israel junction condition. 
In this more general background field configuration, 
we have again found many occasions to exist static bulk spactime 
consistent with the dynamic domain wall. 
In several cases, depending upon the values of the parameters, 
in the early stage of evolution, 
the domain wall is found to have 
an inflationary phase for finite period of it's proper 
time followed by usual decelerated expansion.

\end{abstract}
\section{Introduction}\label{intro}
Our universe as a four dimensional subspace in an extra 
dimensional spacetime, has long been the subject of interest 
from the theoretical as well as phenomenological point of view.
However, with the present day experimental resolution,
we have not seen yet the extra dimensions. This leads to 
a general belief for a long time that if 
there exist extra dimensions then that  
should be compactified to a very
tiny scale (down to a Planck scale). Parallel to this notion of 
compactified dimension, the references  
\cite{joseph,akama,rubakov,other} had come up with a novel
idea of our universe as a domain wall
of spatial dimension $n -2$ in $ n $ dimensional spacetime with uncompactified
extra dimension (more recently \cite{randall}). This means our four dimensional universe is a 
hypersurface moving in an extra spacelike dimension. The primary
assumption of all these models is the localization of standard model
fields on this hypersurface. The localized fields can be
thought of as either, the zero modes of all the bulk fields peaked at the 
position of the domain wall \cite{rubakov,dvali} or by some mechanism, the fields
being polarised parallel to the domain wall world volume. For the 
later case we know that sting theory gives a possible explanation of 
localization by identifying the domain wall as D-brane on which open string
ends \cite{polchinski}. 
As a result of this new idea of a ``domain wall universe,''
many works have explored the notion in the context
of theoretical generalization as well as in various cosmo-
logical and particle phenomenological model building
\cite{dvali,cvetic,pol,lukus}.

Motivated by these ideas of domain wall for the past few years, 
embedding of a four 
dimensional Friedmann-Robertson-Walker
(FRW) universe was considered in generic bulk 
spacetime background with 
cosmological constant and various other fields\cite{kraus,csaki}. 
It is generically true from the Israel junction 
condition\cite{israel} that various fields in the bulk under 
consideration induce energy density with 
different equation of states on the domain wall.
Furthermore, these equation of states appear to be functions of 
various bulk parameters.
So, by tuning these various bulk
parameters in a model under consideration, one can in principle construct
viable cosmology. In certain settings, people have also found
bouncing cosmology which has got much interests in 
the recent time. The unique feature of this bouncing universe 
is nonsingular transition between a
contracting phase of the scale factor of the wall and a following 
expanding stage\cite{sudipto,novello}. However in the recent
studies, people have found some kind of inherent instability
in this bouncig cosmological model showing the very presence
of the singularity \cite{rob}. This also leads to a new direction to study of
circumventing the singularity in the extra dimensional scenario 
\cite{qvedo,risi}.

In this report,  we are not going to construct any 
cosmlogical model. We will first try to generalise
the constructions given in \cite{chamblin}. 
The explicit model for the cosmology, we leave
for our future publications. Before going into the 
motivation of our work, it is ought to
mention that the authors of \cite{chamblin} have discussed 
the dynamics of the domain
wall which is coupled only with a dilaton in the bulk spacetime.  
It was shown by suitably choosing various parameters of the model, 
that a domain wall coupled to a dilaton 
field can be dynamic even within the static bulk spacetime background.
As we stated earlier, all these important aspects came from the 
so called Israel junction conditions across the domain
wall.
The condition tells us the specific relation 
between the extrinsic curvature of the domain wall
and the localized energy momentum tensor
of the wall. The boundary condition can be written as
\bea
\{K_{MN} - K h_{MN}\} = \mu h_{MN}
\eea
where $h_{MN}$ is the induced metric on the domain wall, $K_{MN}$ is the 
extrinsic curvature with its trace $K = h^{MN}K_{MN}$.
Finally the main implication of their study was to 
produce successful inflation on domain wall through
bulk energy induction. At this point
we are not going to elaborate on this. 
In our subsequent analysis, we will be showing their results 
analytically as well as graphically at every stage 
in the appropriate limits.

In the context  of
standard four dimensional cosmology, domain wall had been studied 
extensively. These types of domain walls  were supposed to 
have been produced as a stable topological
defects due to phase transition in the early universe \cite{kibble,vilenkin}.
However, finally it appeared that in the context of 
structure formation the topologically stable domain walls
are not compatible with the recent cosmological observations as opposed to the 
inflationary stage in the early universe\cite{Preskill}.

So, in this report we will be discussing on dynamic domain 
wall solutions in a more 
general background field configuration along the line of \cite{chamblin}.
We consider a general $n$ dimensional action with an usual 
Einstein-Hilbert term and a dilaton field $\phi$ non-minimally 
coupled with a $U(1)$ gauge field $A_{\mu}$. The motivation of 
choosing this kind of
Maxwell-dilaton system is to relate with a more 
fundamental theory specifically string theory.
It is generically true that the low energy limit of any 
string theory turns out to be
the supergravity. So, then by doing suitable 
truncation and dimensional reduction of this 
supergravity action one can get a dilaton-Maxwell system 
having exponential coupling between them. In addition 
to this low energy action,
we also assume the dilaton to be coupled with domain 
wall in the same way as was in \cite{chamblin}. 
Now, for suitable solution ansatz for the bulk 
scalar field, we first analytically find 
five different types of metric solutions under specific 
relation among the various constant parameters in the theory.
At this point, it is important to note that we have taken 
into account the full back-reaction of the
domain wall for the bulk spacetime metric. However, apart from studying 
the properties of these various solutions, we also discuss 
the dynamics of a domain wall in those
bulk backgrounds in the spirit of Chamblin-Reall's paper. 
A topic of particular interest in these kind
of scenario is how inflation occurs on the wall. 
As we have mentioned earlier that generically, the motion of a domain 
wall in a higher dimensional background can be
written as a Hubble expansion equation with various kind of 
positive as well as negative energy
density equation of states. 
As was shown in \cite{chamblin} and also here we will again 
see in a more general setting that for a 
wide range of parameter space of the model under 
consideration, the domain wall indeed inflates in the 
early stage of the evolution followed by 
standard decelerated expansion. The bulk spacetime can also be set 
to a static background for this dynamic domain wall. 
The inflation can either be of exponential or 
power law type depending upon the
kind of bulk solution we are considering. 
In the context of large extra dimensional 
brane world scenario \cite{randall,nima}, there exists
a long list of papers \cite{lukus,kaloper,csaki} 
which have been devoted to study these cosmological aspects. 
One important feature in our model as 
opposed to the general large extra dimensional brane world model
is that it can naturally accommodate the inflation as 
well as decelerated expansion phase of the universe. The 
energy density which drives this inflation 
on the domain wall, strictly come from the bulk.

The paper is organised as follows: In the section \ref{sec1}, we
will start with an action corresponding to a domain wall
moving in Maxwell-dilaton background. After this we explicitly 
write down the equations of motion and its boundary conditions at
the position of the domain wall. In the section \ref{sec2}, taking the
static metric ansatz, we shortly re-state the parametrization of the 
domain wall and the expression for the extrinsic curvature.
From the various components of the extrinsic curvature, 
the consistency condition is derived in order to have 
dynamic domain wall coupled to the scalar field. In section \ref{sec3},
we explicitly solve for the metric and study its structure in grate details
in the various limits of radial co-ordinate. We get five different types of bulk background solutions.
In section \ref{sec4}, we study the dynamics of the domain wall
in those various types of metric backgrounds. The induced metric on the 
domain wall is like FRW cosmological metric. So, naturally, the equation,
describing the dynamics will be a Hubble equation which has been
derived from the Israel junction condition. 
Then following the line of \cite{chamblin}, we again plot the
various forms of the potential encountered by the domain wall and qualitatively
study the dynamics under these potential. 
In some cases we show that the bulk metric becomes time dependent.
Furthermore, in many cases, for finite range of the scale 
factor the domain wall  
inflates for finite period of the proper time followed by the usual decelerated expansion in the static background. 
Finally, in the section \ref{con}, we do some concluding 
remarks and state some futures possible extensions.

\section{Action and Einstein equations} \label{sec1} 
We start with an action of Einstein-Maxwell-Dilaton system in the bulk with
arbitrary dimension $n$ and 
a co-dimension one domain wall coupled with the bulk dilaton field,
\begin{equation} \label{action}
S=\int d^n x\sqrt{-g}\left( \frac 1 2 R ~-~ \frac 1 2 
\pr_{A} \phi \pr^{A} \phi ~-~ V(\phi) ~-~ \frac {\lambda}{2}
e^{- 2 \gamma \phi} F_{AB}F^{AB}\right) ~+~ S_{DW},
\end{equation}
where 
\begin{center}
$S_{DW} ~=~  ~-~ \int d^{n-1} x \sqrt{-h} \left(
\{K\} + \bar{V}(\phi) \right)$,
\end{center}
In the above equations $R$ is the curvature scalar. 
$h_{AB}$ is the induced metric on the 
domain wall. As is clear in the limit $\lambda = 0$, we get back the action studied
in \cite{chamblin}.

Now, corresponding Einstein equations are
\bea
&&R_{AB} = \pr_{A} \phi \pr^A \phi + \frac 2 {n -2} V(\phi) g_{AB}+  \lambda 
e^{- 2 \gamma \phi} \left( 2 F_{AC}F_B^C - \frac {1}{n-2} 
 F_{CD}F^{CD} g_{AB}\right) \\
&&D_C \pr^C \phi - \frac {\pr (\phi)}{\pr \phi} + \lambda 
\gamma e^{- 2 \gamma \phi} F_{AB}F^{AB} = 0 \\
&& D_A\left (e^{- 2 \gamma \phi} F^{AB}\right) = 0 
\eea
where, $D_A$ is co-variant derivative with respect to the bulk 
metric. The boundary conditions at the position of the domain wall are
\bea \label{bond}
&&\{K_{MN}\} = - \frac 1 {n -2} \bar V(\phi) h_{MN} \\
&&\{n^{M}\pr_{M} \phi\} = \frac {\pr \bar{V}(\phi)}{\pr \phi}
\eea
where, $n^M$ is the unit normal to the domain wall.
 The first boundary condition comes
from the Israel junction condition across the wall.

\section{Domain wall and its extrinsic curvature} \label{sec2}
In this section, we will shortly review few steps in calculating
the extrinsic curvature of the domain wall and the boundary conditions for 
various fields across the domain wall
following the paper \cite{chamblin}. 
Once again we will consider reflection symmetry($Z_2$) 
across then wall. So, under this 
symmetry, the above boundary condition Eq.\ref{bond} for the extrinsic
curvature turns out to be
\be
K_{MN} = - \frac 1 {2(n -2)} \bar V(\phi) h_{MN} 
\ee
Our aim is to find out the solution for the dynamic domain wall in a
static bulk background. So, keeping this in mind, we consider the 
static spherically symmetric bulk metric ansatz as
\be \label{metric}
ds^2 = - N(r) dt^2 + \frac 1 {N(r)} dr^2 + R(r)^2 d\Omega_{\kappa}^2
\ee
where, we have taken $ d\Omega_{\kappa}^2$ as the line element 
on a $(n -2)$ dimensional space of constant curvature with the metric
$\tilde {g}_{ij}$. The Ricci curvature of this sub-space is 
$\tilde {R}_{ij} = k (n - 3) \tilde {g}_{ij}$ with $ k \in \{-1,0,1\}$

We want to get spherically symmetric bulk solutions corresponding to a 
homogeneous and isotropic induced metric on the domain wall.
Now, let us parametrize the position of the domain wall by giving $r = r(t)$. Equivalently, 
we can introduce a  new time parameter $\tau$ and specify the functions
\bea
r = r(\tau),~~~~;~~~~ t = t(\tau),~~~~;~~~~ R = R(\tau).
\eea
We choose the domain wall proper time $\tau$ such that the following relation is 
satisfied
\bea
N(r) \left(\frac {dt}{d \tau}\right)^2 - \frac 1 {N(r)} \left( \frac {dr} {d\tau}\right)^2 = 1
\eea
This condition ensures that the induced metric on the wall takes the standard Robertson-Walker form,
\bea
ds_{wall}^2 = - d\tau^2 + R(\tau)^2 d\Omega_{\kappa}^2
\eea
So, the size of our domain wall universe is determined by the radial distance,
$R$, which in turn determines the position of it in the bulk spacetime.

However, the unit normal pointing into $r < r(t)$ and the unit tangent to the 
moving wall read as
\bea
&&n_M = \frac {\sqrt{N}} {\sqrt{N^2 - \dot{r}^2}}(\dot{r}, -1, 0,\dots,0), \\
&&u^{M} = \frac {\sqrt{N}} {\sqrt{N^2 - \dot{r}^2}}(1, \dot{r}, 0,\dots,0)
\eea
respectively. Where, $\dot{r} = \frac {\pr r}{\pr t}$.
Defining these tangent and normal to the domain wall, 
we can readily express the induced
metric on the domain wall and its extrinsic curvature as
\bea
&&h_{MN} = g_{MN} - n_{M} n_{N} \\
&&K_{MN} = h_M^P h_N^Q \nabla_P n_Q
\eea

Now, the expressions for the components of the extrinsic curvature 
by using the bulk metric come out to be
\begin{subequations} \label{excomp1}
\bea 
&&K_{ij} = -  \frac {R'}{R} \frac {N^{3/2}}{\sqrt{N^2 - \dot{r}^2}} h_{ij} \\
&&K_{00} = \frac 1 {\dot r} \frac d {dt} 
\left( \frac {N^{3/2}}{\sqrt{N^2 - \dot{r}^2}}\right) .
\eea
\end{subequations}
By substituting the above expressions Eqs.\ref{excomp1} in the Israel junction condition 
Eq.\ref{bond} we get from ${K_{ij}}$ and $K_{00}$ components
\bea \label{Kij}
&&\frac {R'}{R} = \frac{V(\phi)} {2(n-2)} 
\frac {\sqrt{N^2 - \dot{r}^2}}{N^{3/2}} \\
&&\frac 1 {\dot r} \frac d {dt} \left( \frac {N^{3/2}}{\sqrt{N^2 - \dot{r}^2}}\right) =  \frac{V(\phi)} {2(n-2)},
\eea
which gives us the equations of motion for the domain wall.
'Prime' is derivative with respect the bulk radial coordinate $r$

Now, using the expression for $K_{ij}$ into $K_{00}$, and then integrating
one gets
\bea \label{E1}
R' = C \bar{V}(\phi)
\eea
and again by using the above equation Eq.\ref{E1} into the boundary condition for the
scalar field gives us
\bea \label{E2}
\frac {\pr \phi}{\pr R} = - \frac {n -2} R \frac 1 {\bar V} \frac {\pr \bar V}
{\pr \phi}
\eea
This equation has to hold at every point in the bulk visited by the 
domain wall. So, if the wall visits a range of R, then the above equation
can be solved to yield $\phi$ as a function of R without specifying
the bulk potential. This gives us a consistency condition for the 
dynamic domain wall coupled with the bulk scalar field to exist. In the subsequent section,
we will be using this to find the solution for the metric.

\section{The solutions for bulk metric} \label{sec3}
In this section we will calculate various solutions of the metric assuming
static bulk metric configuration. From the above action Eq.\ref{action} and 
using the metric ansatz Eq.\ref{metric}, one can read out the equations
of motion as
\bea \label{eqs}
&&\frac {R''} {R} = - \frac 1 {n-2} \phi'^2\\
&& \frac 1 {2 R^{n-2}} \left\{N \left(  R^{n-2}\right)'\right\}' -
\frac {k(n-3)(n-2)} {2 R^2} = - V - \frac {2 Q^2 \lambda} {R^{2n -4}} 
e^{2 \gamma \phi} \\
&& \frac {n-2}{4 R^{n-2}} \left( N' R^{n-2}\right)' = - V + 
\frac {(n-3) Q^2 \lambda}{R^{2n -4}} e^{2 \gamma \phi} \\
&&\frac 1 {R^{n-2}}\left( \phi' N R^{n-2}\right)' = 
\frac {\pr V(\phi)}{\pr \phi} + 
\frac {2 \lambda \gamma Q^2}{R^{n-2}} e^{2 \gamma \phi}
\eea
Now, we will employ
the Eqs.(\ref{E1},\ref{E2}) to seek the solution of the 
Einstein equations of motion. So, taking the Liouville
type brane potential
\be
\bar V(\phi) = {\bar V}_0 e^{\alpha \phi},
\ee
one can easily get the solution for the scalar field
without specifying the bulk potential, as well as 
for the radius $R(r)$ of the unit sphere $\Omega_k$ as
\begin{subequations} \label{sol}
\bea 
&&\phi = \phi_0 - \frac {\alpha (n-2)}{\alpha^2(n-2) + 1} log (r),\\
&&R(r) = C {\bar V}_0 e^{\alpha \phi_0} 
r^{\frac 1 {\alpha^2(n-2) + 1}}
\eea
\end{subequations}
where $\phi_0$ and $C$ are the integration constants.
Furthermore, in order to have the solution for the bulk metric, we need to specify the 
dilaton potential $V(\phi)$. So, again we take the same Liouville type bulk potential,
\be
V(\phi) = V_0 e^{\beta \phi}
\ee
where, $V_0$ is constant.
However, in our subsequent analysis, we will  
use the above two expressions Eqs.(\ref{sol}) 
for $R$ and $\phi$ as solutions ansatz with respect to the bulk 
equations of motion. Making use of the bulk potential for the 
scalar field, we find five
types of solutions to the equation of motions Eq.\ref{eqs}. 
At this point it is worth mentioning that, 
for $\lambda =0$ case corresponding to
no electromagnetic field in the bulk \cite{chamblin}, one had
three types of solutions for the bulk metric in compatible with
the dynamic domain wall. However 
in what follows, we will be extensively discussing about
the nature of these various types of solutions and 
subsequently the dynamics of the
domain wall under the same bulk spacetime backgrounds.

{\bf Type-I solution}: When, $\alpha = \beta = \gamma = 0$. We note that the 
bulk and brane potential play the role of cosmological constant 
and brane tenson respectively. So, effectively, the action is a
Einstein-Maxwell system with a bulk cosmological constant and domain
wall with tension.

\begin{figure}
\includegraphics[width=6.0in,height=1.815in]{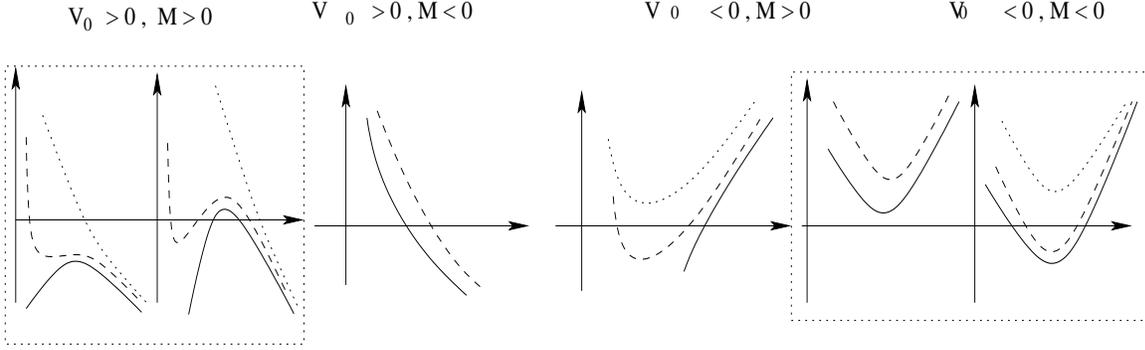}
\caption{N(r) for type-I solution. The solid line indicates
$\lambda = 0$ solution \cite{chamblin}.Dashed and dotted lines
represent the modified solution with $\lambda \ne 0$ for
different sets of parameter values} \label{one}
\end{figure}

By choosing this particular set of value of the parameters, the solution turns out
to be
\bea
&&N(r) = k - 2 M r^{-(n-3)} - \frac {2 V_0}{(n-2)(n-1)}r^2 + 
\frac {2 \lambda Q^2}{(n-3)(n-2)} r^{-2 (n-3)}\\
&&R(r) = r ~~~~;~~~ \phi = \phi_0 
\eea
Where, $M$ and $\phi_0$ are integration constants. 
The solution for the scalar field becomes constant. 
In general, it is difficulty to extract the horizon structure of this
kind of metric solution. So, We have plotted this solution
in the Fig.\ref{one} for several possibilities of parameter values.
Now, it is easy to read off the
horizon structure from these various figures. For all practical purposes 
we have plotted solid lines depicting $\lambda = 0$ case which corresponds to
Einsetin-dilaton system in the bulk \cite{chamblin}.
Whereas, dashed and dotted lines for different values
of the parameters correspond to the full solution
with dilaton-Maxwell fields present in the bulk.
As is seen from the Fig.\ref{one} that for every case, 
there exists singularity at $ r=0$ which is time like.

For all cases, we have four possibilities for different region of the parameter
space ($V_0, M$). When $V_0> 0, M> 0$, one has two different cases, one
for $k = 0, -1$(left panel of the figure) and the another one for $ k = 1$(right panel).
We have noted that for each case the bulk spacetime has horizon. As in the first case
we have  cosmological horizon, but for the second case
there could be a Risner-Nortdstrom(RN) black hole inside
the cosmological horizon \cite{jonson}. In all these
cases asymptotically, the metric becomes de-sitter space
where, $V_0$ is playing the role of cosmological constant.
 
When $ V_0 > 0, M < 0$, the only possibility is a cosmological
horizon with an asymptotically de-sitter space which 
is again defined by the value of $V_0$. So, $r = 0$ is
naked singularity.

If $ V_0 < 0 , M > 0$, for every case $k = 0, \pm 1$, the metric
has either naked singularity at $ r =  0 $ or the same is 
hidden by an event horizon depending upon the values of
various parameters. The black hole is charged in 
an asymptotically anti-de Sitter spacetime \cite{jonson}.
If one takes $r_+$ to be the outer horizon radius, then it
should satisfy
\bea
(n-3) k r_+^{2 n -6} + \frac {2 |{V_0}|}{n -2} r_+^{2 n - 4} - 
\frac {2 \lambda Q^2}{(n-3)(n -2)} \geq 0,
\eea
otherwise there is no horizon. 
When above inequality saturates, the black hole becomes extremal.
The ADM mass $M_{ADM}$ \cite{abbot} of this black hole 
is related to the integration
constant $M$ by
\bea
M_{ADM} = \frac {2 n \omega_{n-1}} 2 M,
\eea
where, $\omega_{n-1}$ is the volume of the unit $n$-sphere.

When $V_0 <0, M< 0$, for $k = 0,1$, the metric has a time like 
naked singularity at $ r= 0$ for any other value of 
the parameters present in the metric as is clear from the expression for
the metric. However, $k = 0$ leads to a possibility
of having a RN black hole in an asymptotically Anti-de Sitter spacetime
for certain range of parameter space $(V_0, M)$ and $ \lambda Q^2$, otherwise it has also naked singularity
at $r = 0$.

\begin{figure}
\includegraphics[width=7.0in,height=4.0in]{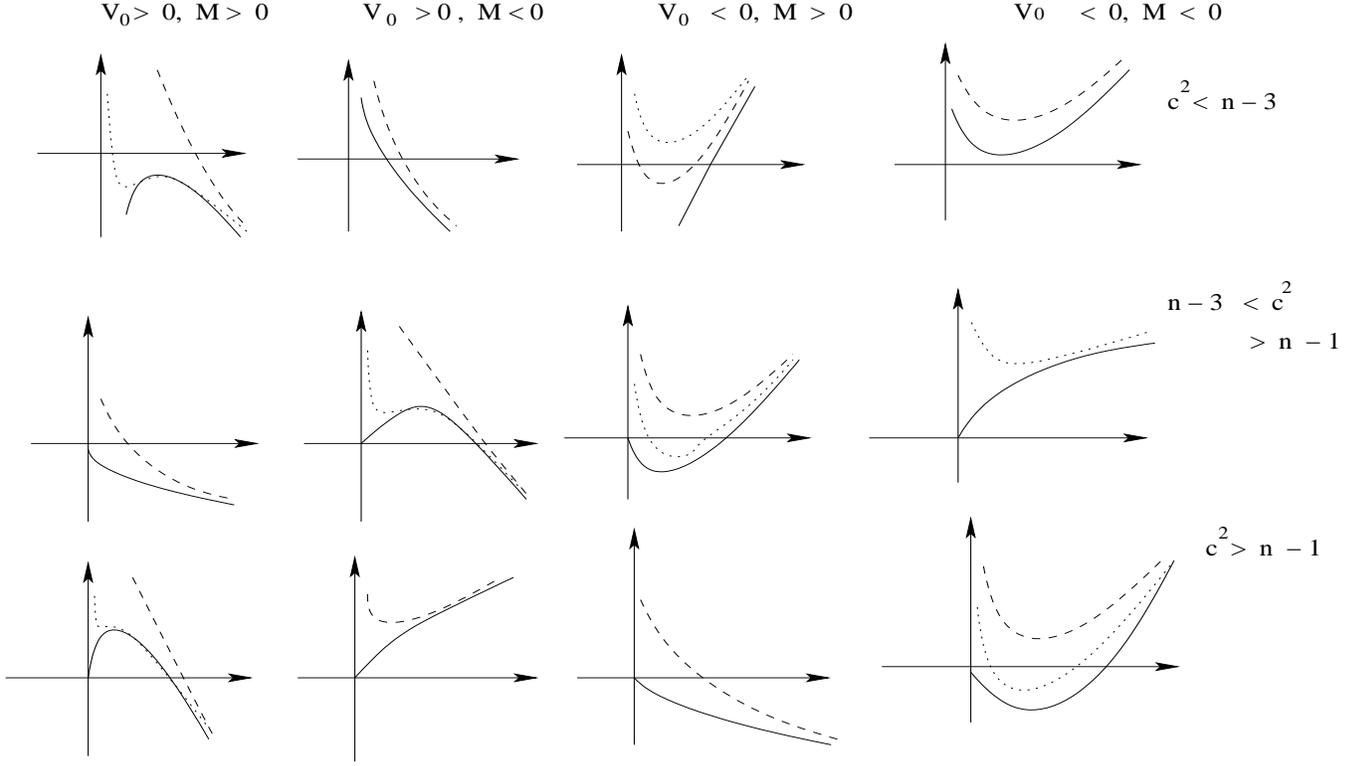}
\caption{N(r) for type-II solutions. $k= 0$} \label{two}
\end{figure}

{\bf Type-II solution}: For $\alpha = \frac {\beta} 2 =  \gamma ~;~ k = 0$, the
bulk metric has only flat spatial section. The 
solution comes out to be
\bea
&&N(r) = -(1 + c^2)^2 r^{\frac 2 {1 + c^2}} \left[ \frac {2 \Lambda}{n - 1 - c^2}
+ 2 M r^{- \frac {n - 1 - c^2}{1 + c^2}} - \frac {2 \lambda \Omega}  
{c^2 + n - 3} r^{- \frac {2(n-2)}{1 + c^2}} \right]\\
&&R(r) = r^{\frac 1 {1 + c^2}} ~~~;~~~ \phi(r) = \sqrt{n-2}\left(
\phi_0* - \frac c {1 + c^2} log (r) \right)
\eea
where $\phi^*_0 = \phi_0/\sqrt{n-2}$, the integration constants. 
The various other notations are given below,
\be
c = \frac 1 2 \beta \sqrt{n-2)}~~~~;~~~~ \Lambda = \frac 
{V_0 e^{2 c \phi^*_0}} {n-2}~~~~;~~~~\Omega = 
\frac {Q^2 e^{2 c \phi^*_0}}{n-2}
\ee
For this type of solution also, we have figured out Fig.\ref{two} the various 
possibilities for different value of the parameters present in the expression for $N(r)$.
This kind of solution had been derived previously in \cite{Cai}
in four dimensional space. In an another work 
\cite{mann}, an explicit solution for the 
metric for $M =0$ has been derived for arbitrary number
of spacetime dimensions. Once again we note that
for this particular choice of parameters,
the Einstein equations of motion are invariant under
constant scale transformation $g_{MN}\rightarrow \omega^2 g_{MN},
\phi \rightarrow \phi- \frac 2{\beta} \log \omega$. 

All the detailed structure can easily be
read off from the corresponding Fig.\ref{two}.
The asymptotic structures remain the same
as was in $\lambda = 0$ case(extensively studied in \cite{chamblin}. 
As is seen from the figures, various structures of the space time are depending 
upon the value of $c$. 
If $c^2 < n-1$, asymptotically for some
range of the parameter values,
we can have either FRW universe ($V_0 > 0$) with flat spatial section
\bea
ds^2 \sim - dT^2 + T^{\frac 2 {c^2}} dx^2
\eea 
by defining $ r $ as time variable $T$ or 
a black $(n-2)$ brane solution for $V_0 < 0$ like
\bea
ds^2 \sim d\rho^2 + \rho^{\frac 2 {c^2}} dx^2.
\eea
by defining $ r $ as space variable $\rho$
Furthermore, if $ c^2 > n-1$, metric
has a curious property that the mass determines the asymptotic
structure. For $M>0$, we have a Kasner type anisotropic cosmological
metric 
\bea
ds^2 \sim - dT^2 + T^{\frac {2(c^2 - n +3)}{ c^2 + n -1}} dt^2 + T^{\frac {4}{c^2 + n -1}} dx^2
\eea
where, the coordinate $r$ is changed 
to time coordinate $T$ and $t$ becomes radial
coordinate. 
 
We have noted that the singularity structures at $r = 0$ for all 
these solutions have drastically changed
due to presence of electromagnetic field and for every case, it is time like. 
When, $V_0 <0, M> 0$, one has
a possibility of having two horizon black hole for $c^2 < n-1$. The same
structure appears again if we take $V_0 <0, M< 0$ and $c^2 > n-1$.
As an alternative behavior, we note that the bulk spacetime has naked singularity
for the above cases.

\begin{figure}
\includegraphics[width=7.0in,height=2.9in]{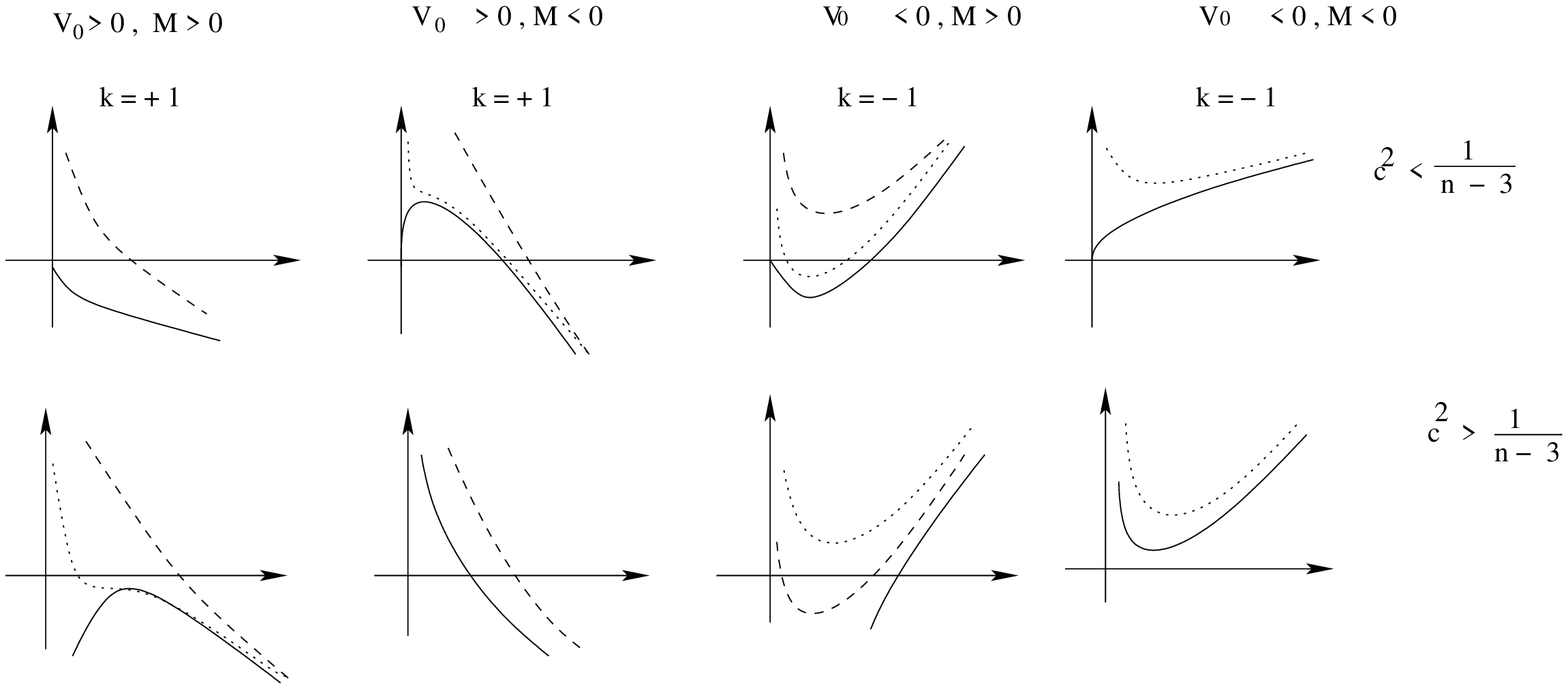}
\caption{N(r) for Type-III solutions. The value of $k$ in the second row
is the same as in the first row when $c^2 < 1$ and opposite of this when $c^2 >1$} \label{three}
\end{figure}

{\bf Type-III solution}: For $\alpha = \frac 2 {\beta(n-2)}  =  \gamma ~;~ k \ne 0$, the
metric has no solution with flat spatial section. The solution looks like
\bea
&&N(r) = - (1 + c^2)^2 r^{\frac 2 {1 + c^2}} \left[ \frac {2 \Lambda}
{(n - 3)c^2 + 1}
+ 2 M r^{- \frac {(n - 3) c^2 + 1}{1 + c^2}} - \frac {2 \lambda \Omega r^{- \frac {2(n-3)c^2 + 2 }{1 + c^2}}}
{\xi^{2n - 4} c^2 \{c^2(n - 3) + 1\}} 
 \right]\\
&&R(r) =\xi r^{\frac {c^2} {1 + c^2}} ~~~;~~~ \phi(r) = \sqrt{n-2}\left(
\phi^*_0 - \frac c {1 + c^2} log r\right)
\eea
where, we use the notation
\bea
\xi = \sqrt{\frac { k (n-3)}{2 \Lambda(1 - c^2)}}
\eea
Again all the solutions are singular at $r = 0$.
The above figure \ref{three} says the detailed asymptotic
structure of the spacetime. 
For this metric, we can analytically solve for the location of the horizon $r = r_h$
where, $N(r_h) = 0$. So, the expression for $r_h$ is
\bea
r_h^{\frac {1 + c^2(n-3)}{1 + c^2}} = - \frac {M (1 + c^2(n-3))} {2 \Lambda}
 \pm \sqrt{ \frac {M^2 (1 + c^2(n-3))^2} {4 \Lambda^2} + 
 \frac {4 \lambda \Omega}{\xi^{2 n -4} c^2 \Lambda}}
\eea
However, as is clear from the above expression and figures that
for $V_0 > 0$, there exists only one horizon. On the other hand,
if we consider $V_0 < 0$ then depending upon the sign of 
parameter $M$ and also value of the various other
parameters, we have either two horizon black hole with open spatial section or a spacetime 
with naked time like singularity at $r = 0$.
As for $V_0< 0$ and $M> 0$, the condition of having the two horizons, among the various parameters
would be look like
\bea
 \frac {M^2 (1 + c^2(n-3))^2} {4 \Lambda^2} -
  \frac {4 \lambda \Omega}{\xi^{2 n -4} c^2 |\Lambda|} \geq 0.
\eea
When, the above inequality saturates then the black hole becomes extremal.

For every case, the asymptotic limit of the solutions depends
upon the sign of the parameter $V_0$ for the bulk scalar field.
This is true even for $\lambda = 0$ which has
extensively been studied in \cite{chamblin}.

{\bf Type-IV solution}: $\alpha = \frac {\beta} 2 
= - \frac{(n-3)}{\gamma (n-2)} ~;~ k = 1$ for $\lambda > 0$ is
considered through out without mentioning further.
So, in this case, the metric with closed spatial section consistent with the dynamic domain wall
is allowed. However, the metric solution looks like 

\bea 
&&N(r) = -(1 + c^2)^2 r^{\frac 2 {1 + c^2}} 
\left[ \frac {2 \Lambda}{n - 1 - c^2}
+ 2 M r^{- \frac {n - 1 - c^2}{1 + c^2}} - \frac {2(n-3) \lambda \Omega}  
{(c^2 + n - 3) c^2 \chi^{2n-4}} r^{ \frac {2( c^2 - 1)}{1 + c^2}}
\right]\\
&&R(r) = \chi r^{\frac 1 {1 + c^2}} ~~~;~~~ \phi(r) = \sqrt{n-2}\left(
\phi^*_0 - \frac c {1 + c^2} log r\right)
\eea
where $M$ and $\phi^*_0$ are the integration constants and
\bea
\chi^{2 n -6} = \frac {2 ((n-3) + c^2)  \lambda \Omega}{k c^2 (n-3)},
\eea
where $c, \Omega$ and $\lambda$ are defined above. 
As stated earlier, it is clear from the 
expression for the $\chi$, that the only possibility could be
$k = 1$ for  $\lambda > 0$. Now, at this point we would like
mention that if we take $\lambda$ to be negative, the energy-momentum tensor
turns out to be that of Kalb-Ramon field \cite{risi} with a different
over all numerical coefficient. We will elaborate on this as a separate 
note at the end. 

Here, also all the solution are singular $r = 0$. Depending
upon the value $c^2$, we have four possibilities.

If $c^2 > 1$, the asymptotic structure of this metric surprisingly
depends on the electric charge $Q^2$ irrespective of the
value of mass parameter $M$ and scalar field potential $V_0$.
By rescaling $t$ and the spatial sections of the metric and changing
the variable $r \rightarrow \rho$, the form of the asymptotic 
metric comes out to be
\bea
ds^2 \sim - \rho^{2 c^2} dt^2 + d\rho^2 + \rho^2 d\Omega_k^2
\eea
for $k =1$. So, the spatial section of
this metric is of cylindrical topology. So, in a large region of parameter
space, we have static bulk metric in the large $R$ limit. 
For few cases, metric has naked singularity but other wise it
is hidden behind black hole horizon. 

On the other hand, when $c^2 < 1$, the singularity behavior at $r =0$ is
characterized by the sing of $V_0$. As is clear from the Fig.4 that
for $V_0 > 0$, the singularity is space-like and vice versa. On the other hand
 the asymptotic structure of the metric is characterized by the sign of 
 $M$. If, $M > 0$ then there are two possibilities. For one $r$ is time 
coordinate every where. In the asymptotic limit, by suitably rescaling the various coordinates,
and changing the variable $r \rightarrow T(r)$ we can write
the metric as
\bea \label{ty4}
ds^2 \sim - dT^2 + T^{\frac 2 {c^2} } dt^2 + T^{\frac 2 {c^2}} d\Omega_k^2
\eea
where, $ k = 1$. So, the metric describes accelerating universe
with spatial sections of cylindrical topology. So, in the 
asymptotic limit, the spatial section of the bulk space time inflates much
faster than that of axial direction. In an another possibility,
we have Schwarzschild-de-Sitter like solution in the bulk but asymptotic structure 
remains the same as Eq.\ref{ty4}.
Whereas if $M < 0$ then $r$ remains spatial coordinate and the asymptotic
solution is the same as the above with the signs of the first
two terms interchanged.
\begin{figure}
\includegraphics[width=7.0in,height=4.50in]{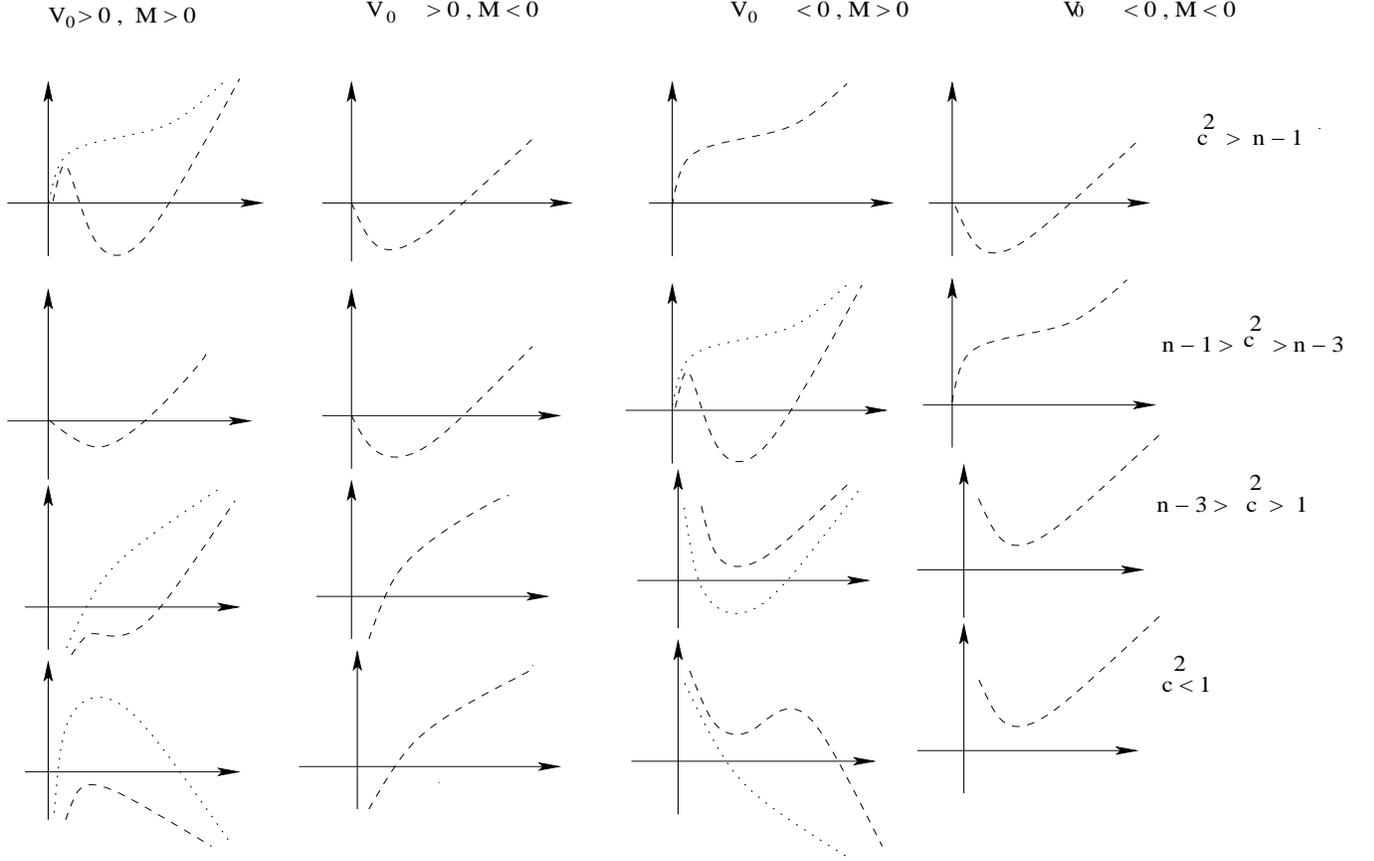}
\caption{N(r) for type-IV solutions. The value of $k = 1$ for $\lambda > 0$} \label{four}
\end{figure}

{\bf Type-V solution}: When $ \alpha = \frac 2 {\beta (n -2)} = - \frac {n-3}{\gamma (n-2)}$, the
metric has again three types of spatial geometry as was in the first case viz. $ k = 0, \pm 1$.

For $k = 0$, the expression for the $N(r)$ turns out to be
\bea \label{first}
&& N(r) = r^2 \left[ M r^{- \frac {(n-2)c^2}{1 + c^2}} + \frac {2 \Lambda (1 + c^2)^2 (n -4)}{(1 + 
c^2(n -3))^2} r^{- \frac {2 c^2}{1 + c^2}} \right] \\
&&R(r) =\eta r^{\frac {c^2} {1 + c^2}} ~~~;~~~ \phi(r) = \sqrt{n-2}\left(
\phi^*_0 - \frac c {1 + c^2} log r\right),
\eea
where, $M$ and $\phi_0^*$ are the integration constants and the expression for $\eta$ is
\bea
\eta^{2n - 4} = - \frac {2 \lambda \Omega (1 + c^2(n -3))}{ 2 \Lambda (1 - c^2)},
\eea
$\Omega$ and $\Lambda$ are already defined earlier. So, 
it is clear from the above expression
for $\eta$ to be positive, either $\Lambda < 0, (1 - c^2) > 0$ or vise versa.

On the other hand when $k \ne 0$, the solution looks like the 
same but the coefficients are different as
\bea \label{second} 
&& N(r) = r^2 \left[ M r^{- \frac {(n-2)c^2}{1 + c^2}} - 2 c^2 (d-3) 
\left(\Lambda + \frac { (n-3) \lambda \Omega}{\zeta^{2n -4}} \right) r^{- \frac {2 c^2}{1 + c^2}} \right] \\
&&R(r) =\zeta r^{\frac {c^2} {1 + c^2}} ~~~;~~~ \phi(r) = \sqrt{n-2}\left(
\phi^*_0 - \frac c {1 + c^2} log r\right),
\eea
where, the expression for $\zeta$ would be the solutions of the following algebraic equation
\bea \label{algeb}
{\cal A} \zeta^{2n-4} - \zeta^{2n-6} + {\cal B} = 0
\eea
where, 
\bea
{\cal A} = \frac {2 \Lambda (1 - c^2)}{k(n-3)}~~~~~;~~~~~ {\cal B} = \frac {2 \lambda \Omega
(1 + c^2(n -3))}{ k (n-3)}
\eea
Now, depending upon the sign of $k, \Lambda$ and $(1 -c^2)$, ${\cal A}, {\cal B}$ could be
positive or negative. So, we have different possibilities which correspond to
different dilaton profile. For example, if ${\cal A} > 0, {\cal B} > 0$ and   
\bea
{\cal A} \left(\frac {n-3}{{\cal A}(n-2)}\right)^{(n-2)} - \left(\frac {n-3}{{\cal A}(n-2)}\right)^{(n-3)}
+ {\cal B} < 0
\eea
above inequality holds then the Eq.\ref{algeb} has two different roots corresponding to
the same value of the parameters we have started with. So, surprisingly, if the metric
has event or cosmological type of horizon, then the same numerical value of the parameters 
give rise to different value of horizon radius and scalar field profile.
For every other possibilities, the Eq.\ref{algeb} has only one positive root.

\begin{figure}
\includegraphics[width=5.40in,height=2.9in]{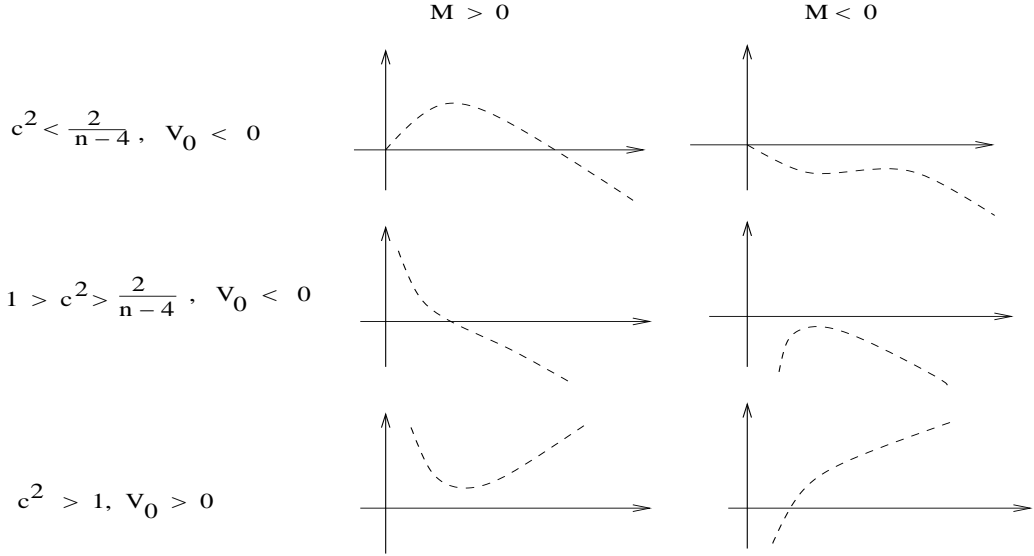}
\caption{N(r) for Type-V solutions} \label{newfive}
\end{figure}

Now, for both the types of metric, one can solve analytically the 
structure of metric function $N(r)$. As is clear from both Eqs.\ref{first},\ref{second} that if 
the metric has either black hole or cosmological type of horizon, the expression for the horizon
radius $r_h$ would be like
\bea
r_h = \left(-\frac M {\cal Z} \right)^{\frac {1 + c^2}{(n-4) c^2}}
\eea
where, either of $M$ and ${\cal Z}$ should be negative. The expression for ${\cal Z}$ is
\bea \label{cond}
{\cal Z} = \left\{\begin{array}{cc}\frac {2 \Lambda (1 + c^2)^2 (n -4)}{(1 + c^2(n -3))^2}&~~~~~~~~\hbox{for}~~~ k = 0\\
 - 2 c^2 (d-3) \left(\Lambda + \frac { (n-3) \lambda \Omega}{\zeta^{2n -4}} \right)&~~~~~~~~
\hbox{for}~~~ k \ne 0 \end{array}\right.
\eea
So, for $k = 0$,  as is also clear from the above Fig.\ref{newfive} as well Eq.\ref{cond}, the metric
has cosmological type of horizon for $M > 0$ and ${\cal Z} < 0$ implying $V_0 < 0$ 
when $1 > c^2 > \frac 2 {n-4}$ and  black hole type
of event horizon for $M < 0$ and $ {\cal Z} > 0$ implying $V_0 > 0$ when $ c^2 > 1$.

Now, for $k \ne 0$, the structure of the spacetime remains the same as for $k =0$, but
in this case for every constraint relations among the parameters, one has four possibilities corresponding
to the value of ${\cal Z}$ in place of $V_0$ and $M$.

It is clear from the metric Eqs.\ref{first},\ref{second} that for any value of $c^2$, the asymptotic
structure of the bulk spacetime is determined by the sign of ${\cal Z}$. For ${\cal Z} > 0$, by changing
$r \rightarrow \rho$ and suitably rescaling the time and space coordinate we have a $n-2$ brane solution
as
\bea
ds^2 \sim d\rho^2 + \rho^{2 c^2}(-dt^2 + d\Omega_k^2) 
\eea
where, the geometry of the spatial section would be any one of $k = 0, \pm 1$.
On the other hand for ${\cal Z} < 0$, the metric will be of FRW cosmological type with
any one of the allowed spatial sections. The scale factor is like $T^{2c^2}$. 

 
So far we have discussed about five possible types of 
solutions for the bulk metric in an
Einstein-Maxwell-Dilaton background. The detailed thermodynamic 
studies of some of
these solutions can be obtained \cite{Cai,mann}. 
In the subsequent section will use
these metrics to study the dynamics of the domain wall. 

\section{Domian wall and its dynamics} \label{sec4}

Without going into further details, we will just state the expression for 
equation of motion of the domain wall as
\bea \label{dome}
\frac 1 2\left( \frac{dR}{d\tau}\right)^2 + F(R) = 0.
\eea
where, $F(R)$ is the expression for the potential encounter by
the domain wall moving in the bulk. The potential is expressed as
\bea
F(R) = \frac 1 2 N R'^2 - \frac 1 {8(n-2)^2} {\bar V}^2 R^2
\eea
where, 'prime' is derivative with respect to bulk radial coordinate $r$.
So, it is clear from the above equation of motion that the 
solution exists only
when $F(R) < 0$. 

Now, in what follows, we consider different types of 
bulk solutions in the expression for the potential of the 
domain wall and study its structure.
We can not solve for the equation of motion analytically for 
its complicated expressions. So, we will try to analyse
the potential on the basis of graphical representation. Furthermore, 
for every case, we will try to display the analytic solutions of the 
evolution equation Eq.\ref{dome} in the asymptotic and near singularity 
region of the bulk spacetime. 

\begin{figure}
\includegraphics[width=6.0in,height=1.815in]{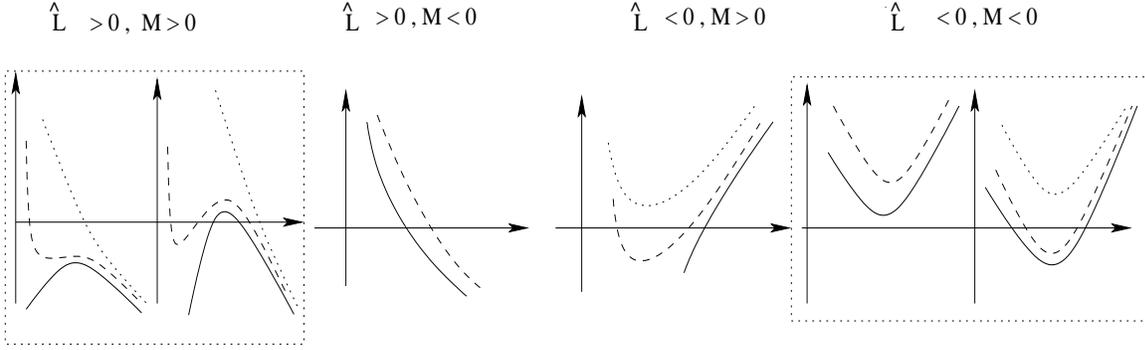}
\caption{F(R) for the type-I solutions.} \label{five}
\end{figure}

{\bf Type-I potential}: The potential encountered 
by the domain wall moving in the bulk spacetime  
for the type-I solution, looks like
\bea
F(R) =  k - 2 M R^{-(n-3)} - \left(\frac {2 V_0}{(n-2)(n-1)} + 
\frac {{\bar V}_0^2}{ 8 (n -2)^2}\right) R^2 +
\frac {2 \lambda Q^2}{(n-3)(n-2)} R^{-2 (n-3)}
\eea
where, the effective cosmological constant on the brane is
\bea
{\hat L} = \frac 1 {n-2} \left[\frac {V_0}{n-1} + 
\frac {{\bar V}^2_0} {8(n-2)}\right]
\eea  
By tuning the bulk and brane potential parameters, we can
set the cosmological constant to be zero. However,
the qualitative features of this potential can be
extracted from the figure \ref{five}. The very fact is that 
for every case, corresponding to a fixed potential
structure, there exists two distinct 
bulk background spacetime depending upon the value of
cosmological constant $V_0$ and the domain wall tension ${\bar V}_0$. 

The plots correspond to these cases are as follows,

 ${\hat L} > 0, M < 0$. In connection with the
choice of parameters, there exist
many possibilities for the structure of the potential.
As we can say that the first figure corresponds
to $ k \ne 1$ and $k = 1$ with specific relation of the parameters.
So, the potential can take 
very distinct form depending on the region of the parameter space. 
The bulk can either be an asymptotically de-sitter
spacetime with single horizon or a topological 
Reissner-Nordstrom(RN) balk hole spacetime. Now in the 
asymptotic limit, the domain wall
goes through an exponential expansion as is clear from the
potential. So, collapsing from the
infinity, the domain wall can either be stopped by the 
repulsive singularity at finite value of $R$ and then re-expands
to infinity, 
or after climbing up the maximum 
of the potential, falls into the local minimum of th potential for 
finite value of $R$ and oscillates. 
If $k = 1$, the background bulk can either be RN-de-Sitter or simply
de-Sitter depending upon the region of
parameter space when $V_0 > 0$. On the other hand for $V_0 < 0$ the bulk
can be either RN-anti-de Sitter or a spacetime which has naked singularity,
with asymptotically anti de-Sitter spacetime. In this case,
one has a possibility of bouncing back the collapsing domain 
wall at finite value of $R$ which is greater than outer
horizon of RN black hole in the bulk. 
In the lower $R$ limit, there exists a region for finite 
value of the scale factor, in which
the domain wall oscillates and in this region domain wall passes through
inflation for finite period followed by standard deceleration.

Asymptotically, 
for any parameter value, the metric becomes anti-de Sitter space.
In the both the cases, the expansion in the large $R$ limit is of 
exponential inflationary type $R(\tau) \sim e^{\sqrt{2 {\hat \Lambda}} \tau}$, 

When ${\hat L} > 0, M < 0$, the asymptotic expansion of the
domain wall world volume is of the same form as in the previous case.
But in this case, for all value of $k$, the structure of the
potential remains more or less the same. So, the motion of the
domain wall has the same behavior like it starts collapsing 
from infinity and then re-expands to infinity by bouncing back from certain 
value of $R$.   

For the other two cases in the parameter space of $({\hat L},M)$,
the structure of the potential is more or less same.
There exists a deep in the potential $F(R)$, in which 
the domain wall oscillates in the region of finite value of the scale factor.
During this course of motion, the domain wall is 
appeared to be residing inside the black hole region. It has again inflation for 
the finite period of time followed by decelerated expansion.
Otherwise there exists no solution for the dynamics.

{\bf Type-II potential}: 
For the type-II solution one gets the expression for the potential like
\bea
F(R) = -R^{2(1 - c^2)}\left[ M R^{-(n - 1 -c^2)} + \hat {\Lambda} - 
\frac {\lambda \Omega}{c^2 + n -3} R^{- {2n - 2}}\right]
\eea
where
\bea
\hat{\Lambda} = \frac {e^{ 2 c \phi^*_0}}{n-2} \left[ \frac {V_0}{n - 1 - c^2}
+ \frac { \bar{V}_0^2}{8(n-2)}\right]
\eea

\begin{figure}
\includegraphics[width=7.0in,height=4.0in]{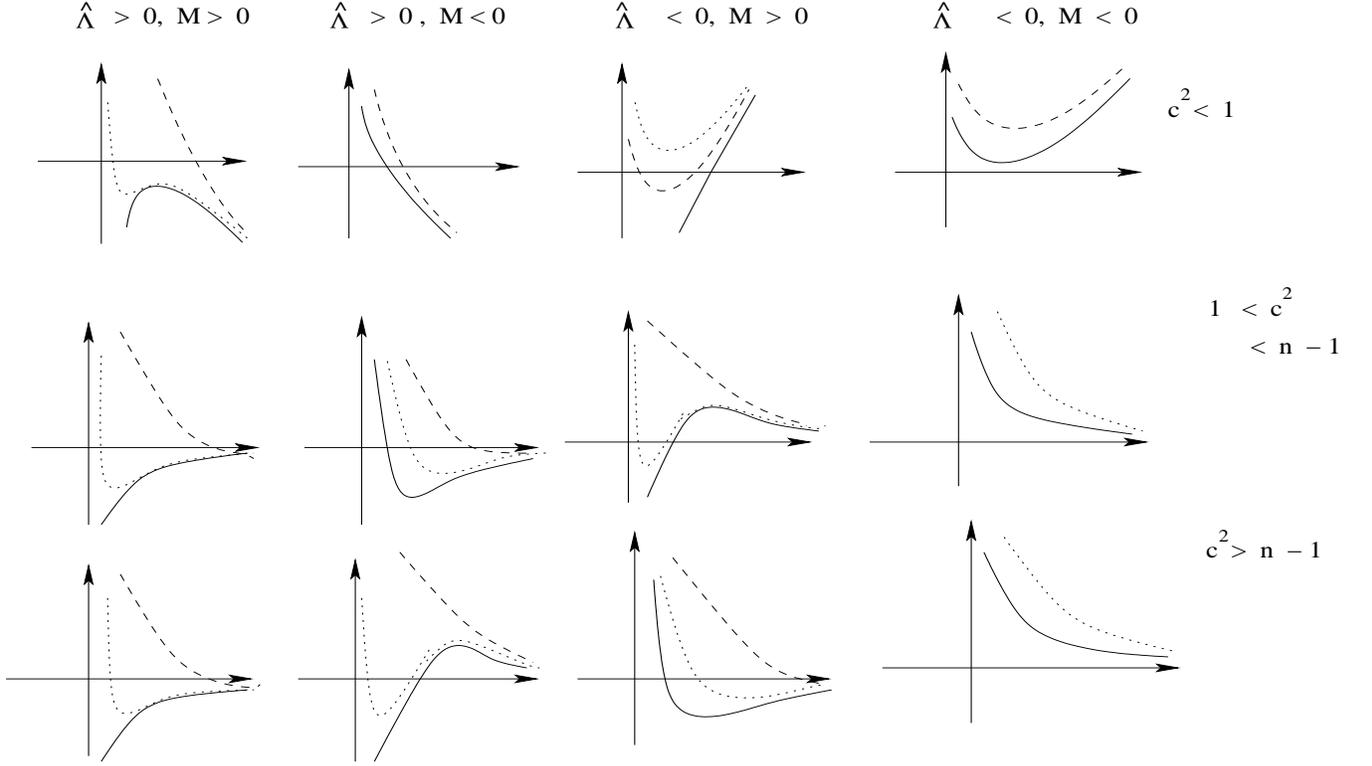}
\caption{$F(R)$ for Type-II solution. $k = 0$} \label{six}
\end{figure}

In this case, the structure of the potential is seen from the
figure \ref{six}. In general, asymptotically the potential function
tends to zero value for $c^2 > 1$ and its form depends upon 
the value of $c^2$ and of course the sign of $M$ and $\hat{\Lambda}$. 
Three classes of behavior are
apparent from the figures,

If the potential function $F(R)$ is positive everywhere. This 
subjects to no solution to the domain wall motion.
For ${\hat \Lambda} < 0$, when $M < 0$, the potential is
positive for all value of $R$ irrespective of the value of $c^2$ 
but when $M> 0$, the same depends upon
constrained region of the full parameter space with $c^2 < n-1$.

As is clear from the figures,for every case 
$F(R)$ is singular but positive in $R\rightarrow 0$ limit.
So, the potential function is always positive for small 
value of $R$. Furthermore if the potential is negative for higher value of $R$ then 
that amounts to a bounce of the domain wall
at finite value of $R$. Asymptotically, the dynamics of 
the domain wall for this type of potential structure, 
is guided by the following respective expressions
\bea
&& R(\tau) = \left( c^2 \sqrt{ 2 \hat{\Lambda}} \right)^{\frac 1 {c^2}} 
\tau^{\frac 1 {c^2}} 
~~~~~~~~~~~~~~~~~~~~~~~~~~~\hbox{when $c^2 < 1$ with $\hat{\Lambda} > 0$} \\
&& R(\tau) = \left( c^2 \sqrt{ 2 \hat{\Lambda}} \right)^{\frac 1 {c^2}} 
\tau^{\frac 1 {c^2}} 
~~~~~~~~~~~~~~~~~~~~~~~~~~~\hbox{when $1 < c^2 < n-1$ with 
$\hat{\Lambda} > 0$},\\
&& R(\tau) = \left( (c^2 + n -2)
\sqrt{ 2 M}\right)^{\frac 1 { (c^2 + n -2)}} 
\tau^{\frac 1 {(c^2 + n -2)}}
~~~~~\hbox{when $c^2 > n-1$ with $ M >0$}.
\eea
The first equation corresponds to inflation. Other two 
correspond to decelerated expansion of the
universe. In the last two cases, namely, $1 < c^2 < n-1 ,
\hat{\Lambda} > 0$ and $c^2 > n-1, M >0$ solutions, we 
have finite period of inflation of the domain wall world volume
in low $R$ limit. In both of these, the domain wall collapses 
from infinity, gets repelled by
the timelike naked singularity in the bulk background, and then expands. 
Inflation occurs when the expansion starts.

If $F(R)$ is negative in the intermediate region of the scale 
factor $R$, there exists two zeros of the potential function. 
This amounts to oscillating as well as bouncing universe. 
The domain wall does not expand to infinity. For
a very particular region of the parameter space, this oscillating 
phase of the universe appears.

{\bf Type-III potential}:
For the type-III solution one gets the expression for the potential like
\bea
F(R) = - \frac {k(n-3)c^4}{2(1-c^2)(1 + c^2(n-3))} &-& M \xi^2 c^4 
\left( \frac R {\xi} \right)^{-\frac { 1 + c^2(n -3)}{c^2}} -
\frac {\bar{V}_0^2 e^{2 \frac {\phi^*_0} {c}} \xi^2} {8(n-2)^2} 
\left( \frac R {\xi} \right)^{-2(\frac 1 {c^2} - 1)} \\
&+& \frac {\lambda \Omega c^2 }{\xi^{2n -6} ( c^2(n-3) + 1)} 
\left( \frac R {\xi} \right)^{-2\frac { 1 + c^2(n -3)}{c^2}}
\eea
In this case also, we can classify the potential into 
four different types of behavior as is seen from Fiq.8.

Class i) $F(R)$ is positive everywhere. Solution does not exist.

Class ii) $F(R)$ is positive for small value of $R$ and 
negative for large value of $R$.
For this class of potential, we have two different behavior 
in the asymptotic limit depending upon the value of $c^2$. 
When $c^2 >1$, asymptotically, the
domain wall is driven by its energy density parametrized by 
${\bar V_0}$ and inflates 
towards infinity under power law of proper time $\tau$ 
\bea \label{inf}
R(\tau) = \left(\frac {{\bar V}_0 e^{\frac {\phi^*_0}{b}}} 
{2 (n-2) c^2} \right)^{c^2} \xi  {\tau^{c^2}},.  
\eea

On the other hand, when $c^2 <1$, the asymptotic dynamics depend
upon the sign of either $k$ or $V_0$. So, when $k = 1$,  
the potential  tends to 
a constant negative value. At late time, the domain wall 
energy density is dominated by the curvature of the spatial 
section and the scale factor of the domain wall grows as
\bea \label{lin}
R(\tau) = \sqrt{\frac{ 2 (n -3) c^4}{(1 - c^2)( 1 + c^2(n -3))}} \tau
\eea
Furthermore, there exists finite amount of inflation in the 
low $R$ limit. The domain wall is repelled by a timelike 
singularity, inflates for a brief period of time and
then decelerates. In an alternative behavior for the some 
what different parameter range, 
the inflationary period does not exists. 

Class iii) $F(R)$ is negative for finite range of $R$. 
This situation occurs when $c^2 < 1$ and $k = -1$ which is 
governed by the sign of $V_0$. The domain
wall world volume describes an open 'oscillating' as well 
as 'bouncing' universe as was explained in
 \cite{chamblin}.

Class iv) $F(R)$ is negative for finite range of $R$ and 
followed by positive value again for finite range of $R$. 
Only some specific range of 
values of the various parameters, we have noted this kind of 
behavior of the potential. For, $c^2 >1$ and $V_0 < 0, M > 0$, 
depending upon the initial position, the domain wall world 
volume can either be describe by a closed 'bouncing' universe or 
a collapsing wall being stopped at some finite 
value of $R$ and again bounced back to infinity. For later case, again
asymptotically, the domain wall inflates according to Eq.\ref{inf}. 

On the other hand, when $c^2 < 1$ and $V_0 > 0, M < 0$, 
qualitatively, the dynamics of the domain wall remains the same 
as above, but in the asymptotic limit the scale factor expands 
linearly with proper time $\tau$ following Eq.\ref{lin}. 	

\begin{figure}
\includegraphics[width=7.0in,height=4.0in]{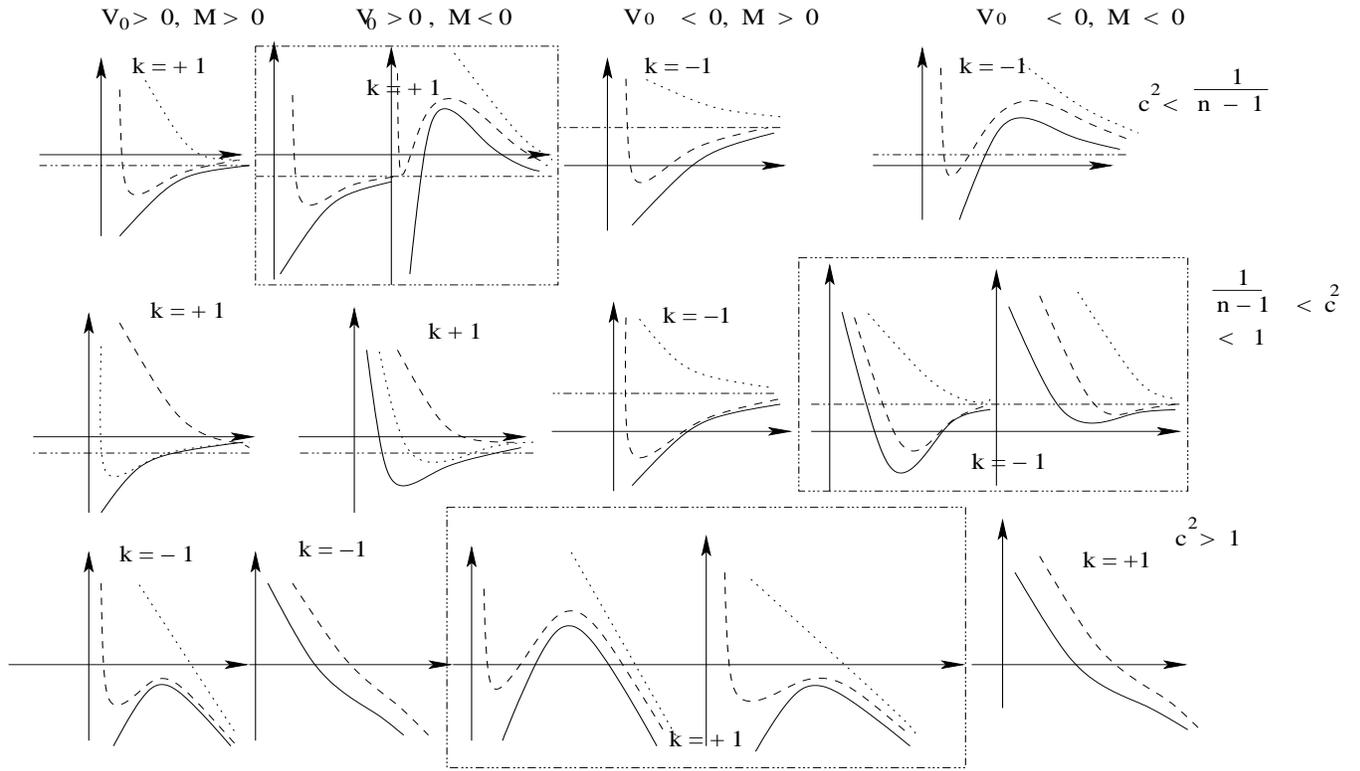}
\caption{$F(R)$ for Type-III solution} \label{seven}
\end{figure}

\begin{figure}
\includegraphics[width=7.0in,height=4.25in]{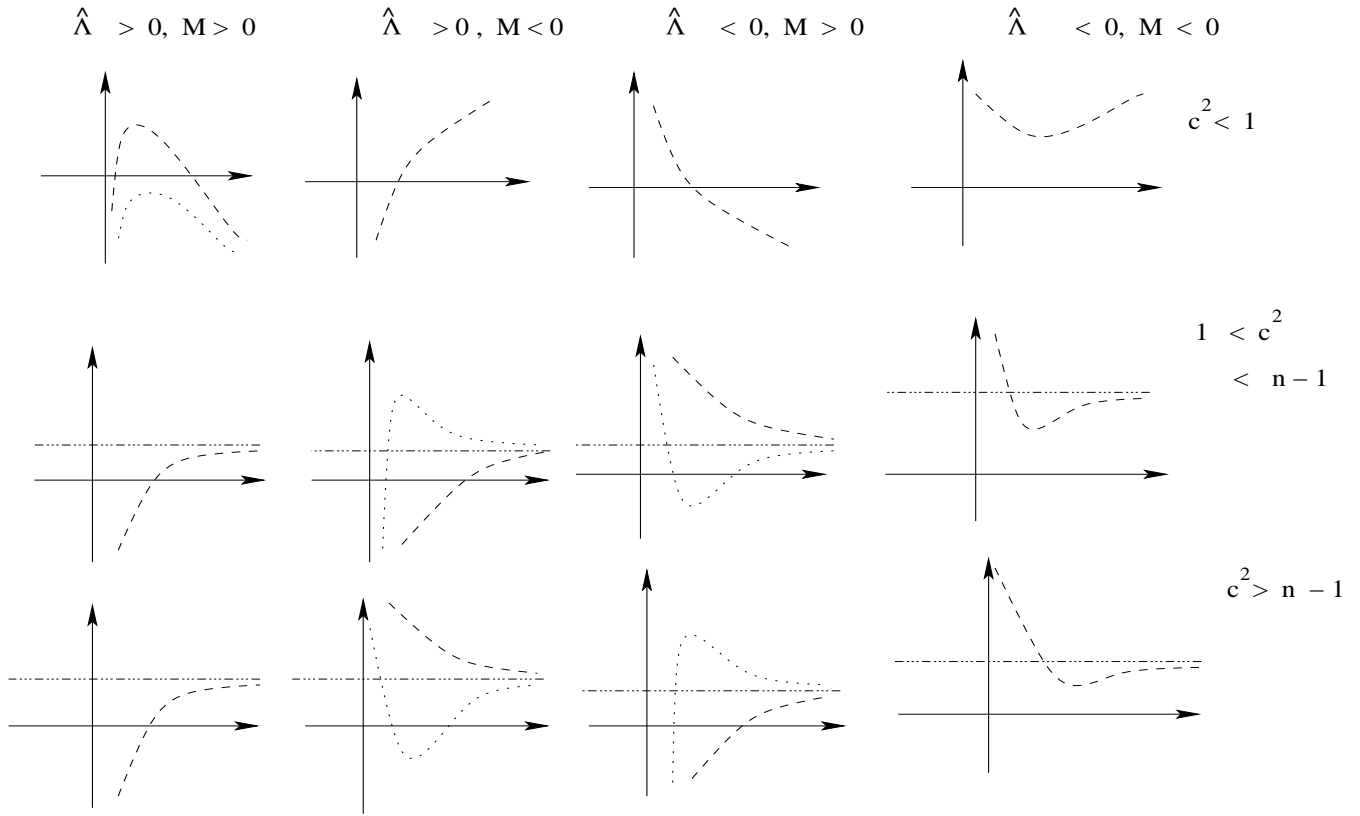}
\caption{$F(R)$ for Type-IV solution.} \label{eight}
\end{figure}

{\bf Type-IV potential}:
For the type-IV solution one gets the expression for the potential like
\bea
F(R) = - \chi^2 \left(\frac {R}{\chi} \right)^{2(1 - c^2)}
\left[ M \left(\frac {R}{\chi} \right)^{-(n - 1 -c^2)} + 
\hat {\Lambda}\right] + 
\frac { k (n-3)^2}{(c^2 + n -3)^2}
\eea
where expression for $\hat{\Lambda}$ is mentioned above.

For $\lambda >0$, as 
we have already mentioned that domain wall has closed spatial 
section. In this case also, we have many different types of 
potential structures corresponding to the values of various 
parameters. Depending upon the value of $c^2$, dynamics is 
determined by $M$ or ${\hat \Lambda}$. From Fig.9, We note  
six different types of structures as follows 

Class i) $F(R)$ is positive everywhere. For 
${\hat \Lambda} < 0, M< 0$, $F(R)$ is positive irrespective 
of the value of $c^2$. As we have mentioned several times that 
we do not have any dynamical
solution of the  domain wall.

Class ii) $F(R)$ is negative everywhere. For a constrained set of
parameters $c^2 <1$ and ${\hat \Lambda} > 0, M> 0$ and as an 
alternative behavior, we get this kind potential. The bulk spacetime may be
either black hole or it has naked singularity at $r = 0$. So, the
domain wall starts collapsing from infinity and falls into bulk 
spacetime singularity. Asymptotic dynamics of it is guided by 
the total brane cosmological constant ${\hat \Lambda}$. So,the 
expression for the scale factor would be like 
\bea \label{type4pot}
R(\tau) = \left( 2 c^4 {\hat \Lambda}\right)^{\frac 2 {c^2}} 
\chi \tau^{\frac 1 {c^2}}.
\eea
which is inflating.  

Class iii) $F(R)$ positive for small value of $R$ but 
negative for large value of $R$. In this case, the domain wall
starts collapsing from infinity and getting repelled by the 
time like singularity at finite value of $R$ re-expands  
again to infinity. The background bulk may have a naked singularity
or a topological black hole with single or double horizon. 
Asymptotic dynamics of the domain
wall is same as Eq.\ref{type4pot}. 

Class iv) $F(R)$ is negative for small value of $R$ but 
positive for large value of $R$. In a large region of the 
parameter space of $({\hat \Lambda}, M)$, the domain
wall encounters this specific potential. So, as Hubble 
equation tells, for almost
all cases, dynamics of the domain wall is confined inside the 
black hole region and is attracted by the singularity at $r = 0$.

Class v) $F(R)$ is positive for finite value of $R$. 
For $c^2 < 1$ and ${\hat \Lambda} > 0, M >0$,
we have this kind of potential structure. Form of the 
${\hat \Lambda}$ suggests that
for the bulk we have either two horizon or a single horizon 
topological black hole. Dynamics are
of two kinds, either it is like class-(iv) for small value 
of the scale factor $R$ or 
in the large $R$, it is like class-(ii). But for the 
later case, asymptotically 
the domain wall inflates following the power low in terms of 
proper time $\tau$ as
$\sim \tau^{\frac 1 {c^2}}$.

Class vi) $F(R)$ is negative for finite range of $R$. 
This kind of potential structures occurs
for ${\hat \Lambda} < 0, M > 0$ with $ 1 <c^2 < n-1$ 
and ${\hat \Lambda} > 0, M < 0$ with $ c^2 > n-1$.
For most of the cases, the bulk back ground has singularity 
hidden by the event horizon and outside
the horizon the spacetime is static. So, again in this case, 
the domain wall has finite period of inflation at low value of $R$ and then 
decelerates and then stopped at some point.
Domain wall describes bouncing universe.

{\bf Type-V potential}:
For type-V solution one gets the expression for the potential like
\bea
F(R) = \frac 1 2 {\Theta}^2  \left(\frac {c^2}{1 + c^2}\right)^2 \left[ M 
\left(\frac {R}{{\Theta}}\right)^{- (n -4)} + {\cal Z} \right] - \frac
{{\bar V}_0^2 e^{2 \alpha \phi_0^*}}  {8 (n-2)^2 { \Theta}^2}
\eea
where ${\cal Z}$ is defined in eq.\ref{cond} and 
\bea
{\Theta} = \left\{\begin{array}{cc} \eta & \hbox{for}~~~ k = 0 \\
                 \zeta & \hbox{for}~~~ k \ne 0
                \end{array} \right.
\eea
Now, as is clear from the above expression for the potential, 
we can solve analytically
the dynamics of the domain wall. The equation we need to solve is 
\bea \label{do1}
\frac {dR}{d \tau} = \sqrt{{\cal D} R^{-(n-4)} + {\cal F}}
\eea
where, 
\bea
{\cal D} = -   \left(\frac {c^2}{1 + c^2}\right)^2  M 
\left(\frac 1 {{\Theta}}\right)^{- (n -6)}~~;~~{\cal F} =\frac
{{\bar V}_0^2 e^{2 \alpha \phi_0^*}}  {4 (n-2)^2 { \Theta}^2}  - {\Theta}^2  \left(\frac {c^2}{1 + c^2}\right)^2 {\cal Z}
\eea
In general for arbitrary dimension $n$, the solution of the above eq.\ref{do1} will be
Hypergeometric function of $R$ like
\bea
R^{\frac {n-2} 2} {_2F_1}\left[\frac {n-2}{2(n-4)}, \frac 1 2, \frac {3n -10}{2(n-4)},
- \frac {\cal F}{\cal D} R^{n-4} \right] = \frac {\sqrt{\cal D}(n -2)} 2 \tau
\eea

 So, it is very difficult to get the inverse of above solution in term of proper time $\tau$. However, 
interestingly, the $R$ dependent part of the potential is seemed to appear from an effective dust like matter
field on the brane. But we note that the 
energy density is a function of bulk electromagnetic charge $Q^2$. 
However, in the early stage of the evaluation, for say $n = 5$, the domain
wall world volume expands like pressure less matter dominated universe
\bea
R(\tau) = 
{{\cal D}^{\frac 1 3}}\left(\frac { 3 \tau}{2} \right)^{\frac 2 3}
\eea
After passing through this matter dominated phase of evolution,
at late time the domain wall world volume expands linearly with
proper time.

\section{Conclusion} \label{con}
To summarize, in this report we have 
tried to generalize  the construction of 
\cite{chamblin} by introducing a $U(1)$ gauge field  
in the bulk. The bulk dilaton field is also  
assumed to couple exponentially with the electromagnetic field 
with an arbitrary coupling parameter $\gamma$. 
Under this some what general background field configuration, 
we have first tried to find out the possible background
solutions taking into account the domain wall back-reaction. 
We have analytically found five different types of solutions 
in accord with the specific relations among the various
parameters. The analytical study of these various metrics 
is very difficult. So, we have adopted the same
line as in \cite{chamblin} by plotting the all
metric functions and studied its structure in various
limits along the radial coordinate. For consistency check, we
have got the same background metric of \cite{chamblin}
under the $\lambda$ tends to 
zero limit for first three solutions. For the other
two cases, we don't have such limits.
Finally, after getting details of the background spacetime
we have tried to study the dynamics of the domain wall in 
those bulk spacetime configurations. 
In this case also, there exist specific relation 
among the various coupling parameters so that one can have a
static bulk spacetime background in consistent with
the dynamic domain wall.

In many cases again, we also found to exist  
inflation for finite period of proper time with
respect to domain wall world volume followed by standard decelerated 
expansion phase. This kind of features might lead to
construct the viable cosmological model in the 
domain wall scenario. One important aspect which we have already
mentioned earlier that in the domain wall
expansion equation which is basically Hubble equation, we 
have encountered the negative energy density. Important
thing with this negative energy density is to lead
to a bouncing cosmology which avoids the bulk singularity for
finite minimum value of the scale factor $R(\tau)$. 
This has already been discussed in many situation only
for the first solution. But we have several solutions with 
different asymptotic as well as near singularity structure
for the same kind of background field configurations. So,
it would be interesting to study in details about the other
solution on this particular context of bouncing cosmology.

Other important attempts would be to interpret  
the various bulk energy density playing the role of different types of 
unseen energy density with respect to domain wall point of view. For
example, this interesting behavior 
may help us to construct dark matter 
and dark energy \cite{dark} model building \cite{supratik} 
in solving discrepancies with standard general relativity predictions 
for the galaxy rotation curves \cite{galaxy}, late time acceleration of the universe \cite{acce},
gravitational lensing \cite{lensing}. 
As an another possible interesting extension of this work 
would be to analyze stability under 
perturbation in the domain wall world volume. 
An interesting point to analyze would be whether all these types of
solutions are compatible in addition to external matter
sources such as radiation and baryonic matter, restricted
to the domain world volume.

\vspace{.1cm}
\noindent
{\bf Acknowledgment}\\
I would like to acknowledge Prof. Soumitra SenGupta for carefully
reading the manuscript and also for valuable suggestions.
I also thank Prof. Sudipta Mukherji for his valuable 
comments during the course of this work.  

{\bf Note added}: At this stage we would like to illustrate a point 
regarding a recent study \cite{risi} in which author has found
the solution for the bouncing universe by using the Kalb-Ramnod(KR) (second rank antisymmetric
tensor $B_{MN}$) field in the static bulk. In doing so, the author has
chosen a particular solution ansatz for the KR field. We can reproduce that 
results just by changing the sign of the parameter $\lambda$ in our
action with an appropriate numerical coefficient. We take Kalb-Ramond 
field in stead of Maxwell field in the bulk like $F_{MN}F^{MN}
\rightarrow H_{MNP}H^{MNP}$ where $H_{MNP} = \partial_{[M}B_{NP]}$ 
being strength of the
$B_{MN}$. After getting equations of motion, 
we make a particular solution ansatz 
(only for $n = 5$ dimension) like $H^{MNP} = e^{2 \gamma \phi} \epsilon^{MNPQR} \partial_{Q} C_R$ where $C_M$ is some dual vector field 
resembling to the electromagnetic potential. 
This particular ansatz leads us to the solutions which we have 
discussed through out this report, with $(\lambda)$ to be replaced by 
$(- \lambda)$ up to a numerical factor depending upon the rank of the KR field.
The structure of the bulk spacetime solutions get reversed at $r =0$ for the first three types of solutions
and corresponding potentials. For the fourth type, the from of the solutions remains the
same but in this case, geometry of the spatial section for the bulk spacetime will be
open ($ k = -1$) as opposed to the electromagnetic case. 
The structure of all the solutions for the finite range of $R$ will get
changed drastically.

\end{document}